\documentstyle[12pt]{article}
\def\be {\begin{equation}}
\def\ee {\end{equation}}
\def\ba {\begin{eqnarray}}
\def\ea {\end{eqnarray}}
\newcommand{\bq}{\begin{eqnarray}}
\newcommand{\eq}{\end{eqnarray}}

\def\bi {\begin{itemize}}
\def\ei {\end{itemize}}
\begin{document}
\def\bea{\begin{eqnarray}}
\def\eea{\end{eqnarray}}
\title{\bf {Interacting generalized Chaplygin gas model in non-flat universe}}
 \author{M.R. Setare  \footnote{E-mail: rezakord@ipm.ir}
  \\ {Department of Science,  Payame Noor University. Bijar, Iran}}
\date{\small{}}
\maketitle
\begin{abstract}
We employ
 the generalized Chaplygin gas of interacting dark energy to obtain the equation of state
   for the generalized Chaplygin gas energy density
 in non-flat universe. By choosing a negative value for $B$ we see that $w_{\rm
\Lambda}^{eff}< -1$, that corresponds to a universe dominated by
phantom dark energy.
 \end{abstract}

\newpage
\section{Introduction}
One of the most important problems of cosmology, is the problem of
so-called dark energy (DE). The type Ia supernova observations
suggests that the universe is dominated by dark energy with negative
pressure which provides the dynamical mechanism of the accelerating
expansion of the universe \cite{{per},{gar},{ries}}. The strength of
this acceleration is presently matter of debate, mainly because it
depends on the theoretical model implied when interpreting the data.
Most of these models are based on dynamics of a scalar or
multi-scalar fields. Primary scalar field candidate for dark energy
was quintessence scenario\cite{{rat},{zlat}}, a fluid with the
parameter of the equation of state lying in the range, $-1< w< {-1
\over 3}$. While the most model independent analysis suggest that
the acceleration of the universe to be below the de Sitter value
\cite{Daly}, it is certainly true that the body of observational
data allows for a wide parameter space compatible with an
acceleration larger than the de Sitter's \cite{{cal},{hans}}. If
eventually this proves to be the case, the fluid driving the
expansion would violate not only the strong energy condition $\rho +
3P>0$, but the dominate energy condition $\rho + P>0$, as well.
Fluids of such characteristic dubbed phantom fluid \cite{cal1}. In
spite of the fact that the field theory of phantom fields encounter
the problem of stability which one could try to bypass by assuming
them to be effective fields \cite{{car},{gib}}, it is nevertheless
interesting to study their cosmological implication. Recently there
are many relevant studies on phantom energy \cite{meng}. The
analysis of the properties of dark energy from recent observations
mildly favor models with $w$ crossing -1 in the near past. So far, a
large class of scalar-field dark energy models have been studied,
including quintessence, K-essence \cite{kessence}, tachyon
\cite{tachyon}, ghost condensate \cite{ghost1,ghost2} and quintom
\cite{quintom}, and so forth.  In addition, other proposals on dark
energy include interacting dark energy models \cite{intde},
braneworld models \cite{brane}, and holographic dark energy models
\cite{holo}, etc..
\\
 In a
very interesting paper Kamenshchik, Moschella, and Pasquier
\cite{kmp}have studied a homogeneous model based on a single fluid
obeying the Chaplygin gas equation of state \be \label{chp}
P=\frac{-A}{\rho} \ee where $P$ and $\rho$ are respectively pressure
and energy density in comoving reference frame, with $\rho> 0$; $A$
is a positive constant. This equation of state has raised  a certain
interest \cite{jac} because of its many interesting and, in some
sense, intriguingly unique features. Some possible motivations for
this model from the field theory points of view are investigated in
\cite{a}. The Chaplygin gas emerges as an effective fluid associated
with d-branes \cite{b} and can also be obtained from the Born-Infeld
action \cite{c}.\\
Inserting the equation of state (\ref{chp}) into the relativistic
energy conservation equation, leads to a density evolving as \be
\label{enerd}\rho_{\Lambda}=\sqrt{A+\frac{B}{a^{6}}} \ee where $B$
is an integration constant.\\
In present paper, using the generalized Chaplygin gas model of dark
energy, we obtain equation of state for interacting Chaplygin gas
energy density in non-flat universe. The current available
observational data imply that the dark energy behaves as
phantom-type dark energy, i.e. the equation-of-state of dark energy
crosses the cosmological-constant boundary $w=-1$ during the
evolution history. We show this phantomic description of the
interacting generalized Chaplygin gas dark energy in non-flat
universe with $B<0$.
\section{ Interacting generalized Chaplygin gas }
In this section we obtain the equation of state for the generalized
Chaplygin gas when there is an interaction between generalized
Chaplygin gas energy density $\rho_{\Lambda}$ and a Cold Dark
Matter(CDM) with $w_{m}=0$. The continuity equations for dark energy
and CDM are
\begin{eqnarray}
\label{2eq1}&& \dot{\rho}_{\rm \Lambda}+3H(1+w_{\rm \Lambda})\rho_{\rm \Lambda} =-Q, \\
\label{2eq2}&& \dot{\rho}_{\rm m}+3H\rho_{\rm m}=Q.
\end{eqnarray}
The interaction is given by the quantity $Q=\Gamma \rho_{\Lambda}$.
This is a decaying of the generalized Chaplygin gas component into
CDM with the decay rate $\Gamma$. Taking a ratio of two energy
densities as $r=\rho_{\rm m}/\rho_{\rm \Lambda}$, the above
equations lead to
\begin{equation}
\label{2eq3} \dot{r}=3Hr\Big[w_{\rm \Lambda}+
\frac{1+r}{r}\frac{\Gamma}{3H}\Big]
\end{equation}
 Following Ref.\cite{Kim:2005at},
if we define
\begin{eqnarray}\label{eff}
w_\Lambda ^{\rm eff}=w_\Lambda+{{\Gamma}\over {3H}}\;, \qquad w_m
^{\rm eff}=-{1\over r}{{\Gamma}\over {3H}}\;.
\end{eqnarray}
Then, the continuity equations can be written in their standard form
\begin{equation}
\dot{\rho}_\Lambda + 3H(1+w_\Lambda^{\rm eff})\rho_\Lambda =
0\;,\label{definew1}
\end{equation}
\begin{equation}
\dot{\rho}_m + 3H(1+w_m^{\rm eff})\rho_m = 0\; \label{definew2}
\end{equation}
We consider the non-flat Friedmann-Robertson-Walker universe with
line element
 \be\label{metr}
ds^{2}=-dt^{2}+a^{2}(t)(\frac{dr^2}{1-kr^2}+r^2d\Omega^{2}).
 \ee
where $k$ denotes the curvature of space k=0,1,-1 for flat, closed
and open universe respectively. A closed universe with a small
positive curvature ($\Omega_k\sim 0.01$) is compatible with
observations \cite{ {wmap}, {ws}}. We use the Friedmann equation to
relate the curvature of the universe to the energy density. The
first Friedmann equation is given by
\begin{equation}
\label{2eq7} H^2+\frac{k}{a^2}=\frac{1}{3M^2_p}\Big[
 \rho_{\rm \Lambda}+\rho_{\rm m}\Big].
\end{equation}
Define as usual
\begin{equation} \label{2eq9} \Omega_{\rm
m}=\frac{\rho_{m}}{\rho_{cr}}=\frac{ \rho_{\rm
m}}{3M_p^2H^2},\hspace{1cm}\Omega_{\rm
\Lambda}=\frac{\rho_{\Lambda}}{\rho_{cr}}=\frac{ \rho_{\rm
\Lambda}}{3M^2_pH^2},\hspace{1cm}\Omega_{k}=\frac{k}{a^2H^2}
\end{equation}
Now we can rewrite the first Friedmann equation as
\begin{equation} \label{2eq10} \Omega_{\rm m}+\Omega_{\rm
\Lambda}=1+\Omega_{k}.
\end{equation}
Using Eqs.(\ref{2eq9},\ref{2eq10}) we obtain following relation for
ratio of energy densities $r$ as
\begin{equation}\label{ratio}
r=\frac{1+\Omega_{k}-\Omega_{\Lambda}}{\Omega_{\Lambda}}
\end{equation}
In the generalized Chaplygin gas approach \cite{c}, the equation of
state to (\ref{chp}) is generalized to \be \label{chpge}
P_{\Lambda}=\frac{-A}{\rho_{\Lambda}^{\alpha}} \ee The above
equation of state leads to a density evolution as \be \label{denge}
\rho_{\Lambda}=[A+\frac{B}{a^{3(1+\alpha)}}]^{\frac{1}{1+\alpha}}
\ee Taking derivative in both sides of above equation with respect
to cosmic time, we obtain \be \label{timde}\dot{\rho_{\Lambda}}=-3B
H a^{-3(1+\alpha)}[A+B a^{-3(1+\alpha)}]^{\frac{-\alpha}{1+\alpha}}
\ee Substituting this relation into Eq.(\ref{2eq1}) and using
definition $Q=\Gamma \rho_{\Lambda}$, we obtain
\begin{equation}\label{stateq}
w_{\rm \Lambda}=\frac{B}{a^{3(1+\alpha)}[A+B
a^{-3(1+\alpha)}]}-\frac{\Gamma}{3H}-1.
\end{equation}
Here as in Ref.\cite{WGA}, we choose the following relation for
decay rate
\begin{equation}\label{decayeq}
\Gamma=3b^2(1+r)H
\end{equation}
with  the coupling constant $b^2$. Using Eq.(\ref{ratio}), the above
decay rate take following form
\begin{equation}\label{decayeq2}
\Gamma=3b^2H\frac{(1+\Omega_{k})}{\Omega_{\Lambda}}
\end{equation}
Substituting this relation into Eq.(\ref{stateq}), one finds the
generalized Chaplygin gas energy equation of state
\begin{equation}\label{stateq1}
w_{\rm \Lambda}=\frac{B}{a^{3(1+\alpha)}[A+B
a^{-3(1+\alpha)}]}-\frac{b^2(1+\Omega_{k})}{\Omega_{\rm \Lambda}}-1.
\end{equation}
Now using the  definition generalized Chaplygin gas energy density
$\rho_{\rm \Lambda}$, and using $\Omega_{\Lambda}$, we can rewrite
the above equation as
\begin{equation}\label{stateq2}
w_{\rm \Lambda}=\frac{B}{(3M_p^2H^2 a^3 \Omega_{\rm
\Lambda})^{1+\alpha}}-\frac{b^2(1+\Omega_{k})}{\Omega_{\rm
\Lambda}}-1.
\end{equation}
 From
Eqs.(\ref{eff}, \ref{decayeq2}, \ref{stateq2}), we have the
effective equation of state as
\begin{equation} \label{3eq401}
w_{\rm \Lambda}^{eff}=\frac{B}{(3M_p^2H^2 a^3 \Omega_{\rm
\Lambda})^{1+\alpha}}-1.
\end{equation}
By choosing a negative value for $B$ we see that $w_{\rm
\Lambda}^{eff}< -1$, that corresponds to a universe dominated by
phantom dark energy. Eq.(\ref{3eq401}), for $\alpha=1$, is the
effective parameter of state for Chaplygin gas. In this case, in the
expression for energy density (\ref{enerd}), term under square root
should be positive, i.e. $a^6>\frac{-B}{A}$, then the minimal value
of the scale factor is given by
$a_{min}=(\frac{-B}{A})^{\frac{1}{6}}$, therefore according to this
model we have a bouncing universe. Generally for this model $A > 0$,
$B < 0$ and $1 + \alpha
 > 0$. From Eq. (\ref{denge}
), we can realize that the cosmic scale factor takes values in the
interval $a_{ min }< a <\infty$ which corresponds to $0 < \rho
<(2A)^{\frac{1}{1+\alpha}}$, where

\be \label{lagp} a_{ min }=
(\frac{-B}{A})^{\frac{1}{3(1+\alpha)}}\ee
 Using Eq.(\ref{enerd}), one can see that the  Chaplygin gas
model interpolates between dust at small $a$ and a cosmological
constant at large  $a$, but choosing a negative value for $B$, this
quartessence idea lose.
 Following \cite
{kmp} if we consider a homogeneous scalar filed $\phi(t)$ and a
potential $V(\phi)$ to describe the Chaplygin cosmology, we find \be
\label{phidot}\dot{\phi}^{2}=\frac{B}{a^{6}\sqrt{A+\frac{B}{a^{6}}}}
\ee Now, by choosing a negative value for $B$ we see that
$\dot{\phi}^{2}<0$, then we can write \be \label{sieq}\phi=i\psi \ee
In this case the Lagrangian of scalar field $\phi(t)$ can rewritten
as \be \label{lag} L=\frac{1}{2}
\dot{\phi}^{2}-V(\phi)=-\frac{1}{2}\dot{\psi}^{2}-V(i\psi)\ee The
energy density and the pressure corresponding to the scalar field
$\psi$ are as following respectively \be
\label{rheq}\rho_{\psi}=-\frac{1}{2}\dot{\psi}^{2}+V(i\psi) \ee \be
\label{preq}P_{\psi}=-\frac{1}{2}\dot{\psi}^{2}-V(i\psi) \ee
therefore, the scalar field $\psi$ is a phantom field. This implies
that one can generate phantom-like equation of state from an
interacting generalized Chaplygin gas dark energy model in non-flat
universe.
\section{Conclusions}
In order to solve cosmological problems and because the lack of our
knowledge, for instance to determine what could be the best
candidate for DE to explain the accelerated expansion of universe,
the cosmologists try to approach to best results as precise as they
can by considering all the possibilities they have. Within the
different candidates to play the role of the dark energy, the
Chaplygin gas, has emerged as a possible unification of dark matter
and dark energy, since its cosmological evolution is similar to an
initial dust like matter and a cosmological constant for late times.
Inspired by the fact that the Chaplygin gas possesses a negative
pressure, people \cite{mas} have undertaken the simple task of
studying a FRW cosmology of a universe filled with this type of
fluid.\\
In this paper, by considering an interaction between generalized
Chaplygin gas energy density and CDM, we have obtained the equation
of state for the interacting  generalized Chaplygin gas energy
density in the non-flat universe. Then we have shown that the
interacting generalized Chaplygin gas dark energy with $B<0$  can be
described
 by the phantom field. Previously I have shown that the phantom dark energy model can behave as
a holographic dark energy \cite{SET}. In the other hand recently
Zimdhal has shown that a Chaplygin gas, can be seen as special
realization of a holographic dark energy cosmology if the option of
an interaction between pressureless dark matter and dark energy is
taken seriously \cite{ZIM}. In fact pressureless dark matter in
interaction with holographic dark energy is more than just another
model to describe an accelerated expansion of the universe. This
clarify the role of interaction in this model.

\end{document}